\documentclass[pra, aps, twocolumn,superscriptaddress, nofootinbib, tightenlines, nobibnotes,floatfix,longbibliography]{revtex4-1}
\usepackage{amssymb}
\usepackage{graphicx}
\usepackage{dcolumn}
\usepackage{bm}
\usepackage{hyperref}
\usepackage[dvipsnames]{xcolor}
\usepackage[caption=false]{subfig}
\usepackage[utf8]{inputenc}
\usepackage[english]{babel}
\usepackage{amsmath}
\usepackage{hyperref}
\usepackage{amsmath}
\usepackage{mathtools} 

\definecolor{lapislazuli}{rgb}{0, 0, 1}
\definecolor{YKblue}{rgb}{0.0, 0.18, 0.65}
\definecolor{carmine}{rgb}{0.81, 0.09, 0.03}
\definecolor{lavender}{rgb}{0.84, 0.49, 0.87}

\hypersetup{
	colorlinks=true,
	linkcolor=lapislazuli,
	linktoc=page,
	citecolor=lapislazuli,
	urlcolor=lapislazuli
}

\usepackage{amssymb}
\usepackage{mathtools}

\everymath{\displaystyle} 
\newcommand{\pr}[1]{\ensuremath{\left[#1\right]}} 
\newcommand{\pc}[1]{\ensuremath{\left(#1\right)}} 
\newcommand{\px}[1]{\ensuremath{\left\lbrace#1\right\rbrace}} 
\newcommand{\av}[1]{\ensuremath{\left\langle#1\right\rangle}} 

\newcommand{\vect}[1]{\boldsymbol{#1}}

\begin{document}
	
	\title{Sub-nK thermometry of an interacting $d$-dimensional homogeneous Bose gas }
	
	\author{Muhammad Miskeen Khan}
	\affiliation{Instituto Superior T\'ecnico, Universidade de Lisboa, Portugal}
	\affiliation{ICFO -- Institut de Ciències Fotòniques, The Barcelona Institute
		of Science and Technology, Av. Carl Friedrich Gauss 3, 08860 Castelldefels (Barcelona), Spain}
	\affiliation{Instituto de Plasmas e Fus\~ao Nuclear, Instituto Superior T\'ecnico, Universidade de Lisboa, Portugal}

	\author{Mohammad Mehboudi}
	\affiliation{D\'epartement de Physique Appliq\'ee, Universit\'e de Gen\`eve, CH-1211 Gen\`eve, Switzerland}
	\affiliation{ICFO -- Institut de Ciències Fotòniques, The Barcelona Institute
		of Science and Technology, Av. Carl Friedrich Gauss 3, 08860 Castelldefels (Barcelona), Spain}
	
	\author{Hugo Ter\c{c}as}
	\affiliation{Instituto Superior T\'ecnico, Universidade de Lisboa, Portugal}
	\affiliation{Instituto de Plasmas e Fus\~ao Nuclear, Instituto Superior T\'ecnico, Universidade de Lisboa, Portugal}
	
	\author{Maciej Lewenstein}
	\affiliation{ICFO -- Institut de Ciències Fotòniques, The Barcelona Institute
		of Science and Technology, Av. Carl Friedrich Gauss 3, 08860 Castelldefels (Barcelona), Spain}
	\affiliation{ICREA, Pg. Llu\'is Companys 23, 08010 Barcelona, Spain}
	
	\author{Miguel Angel Garcia-March}
	\affiliation{ICFO -- Institut de Ciències Fotòniques, The Barcelona Institute
		of Science and Technology, Av. Carl Friedrich Gauss 3, 08860 Castelldefels (Barcelona), Spain}
	\affiliation{Instituto Universitario de Matem\'atica Pura y Aplicada, Universitat Polit\`ecnica de Val\`encia, E-46022 Val\`encia, Spain}

	\begin{abstract}
We propose experimentally feasible means for non-destructive thermometry of {\it homogeneous} Bose Einstein condensates in different spatial dimensions ($d\in\{1,2,3\}$). Our impurity based protocol suggests that the fundamental error bound on thermometry at the sub nano Kelvin domain depends highly on the dimension, in that the higher the dimension the better the precision. Furthermore, sub-optimal  thermometry of the condensates by using measurements that are experimentally feasible is explored. We specifically focus on measuring position and momentum of the impurity that belong to the family of Gaussian measurements. We show that, generally, experimentally feasible measurements are far from optimal, except in 1D, where position measurements are indeed optimal. This makes realistic experiments perform very well at few nano Kelvin temperatures for all dimensions, and at sub nano Kelvin temperatures in the one dimensional scenario. These results take a significant step towards experimental realisation of probe-based quantum thermometry of Bose Einstein condensates, as it deals with them in one, two and three dimensions and uses feasible measurements applicable in current experimental setups.
\end{abstract}
   \maketitle
	\section{Introduction $-$} 
The fundamental and technological importance of temperature of quantum systems has led to the rapid development of the theory of \textit{quantum thermometry}~\cite{Mehboudi2019,de2018quantum}. Among the fundamental questions of interest in this subject are (i) what are the ultimate limits on thermometry precision? and (ii) what is the best measurement?---which is mainly relevant for quantum systems due to the \textit{incompatibility} of measurements.
	Often, quantum thermometry proposals are probe-based, in that, the information about the unknown temperature of a sample is registered on the quantum state of a probe (thermometer) through some sort of \textit{sample-probe} interaction~\cite{PhysRevLett.122.030403,razavian2019quantum,PhysRevLett.123.180602}.
	By performing a suitable measurement on the probe, one aims to infer the temperature with minimal \textit{statistical error}. This forms a non-destructive thermometry strategy, thanks to the fact that the probe size is significantly smaller than the sample, and that the sample is not directly measured. The theory of probe-based thermometry has been adapted to describe various platforms such as Fermionic or Bosonic samples, and by using different resources like entanglement, criticality and coherence~\cite{PhysRevLett.114.220405,PhysRevLett.125.080402,de2016local,Paris_2015,PhysRevLett.122.030403,razavian2019quantum,PhysRevB.98.045101,Mehboudi_2015,PhysRevA.103.023317,PhysRevLett.119.090603,PhysRevLett.123.180602,PhysRevA.95.022121,Campbell_2017,PhysRevLett.125.080402,mok2021optimal,Latune_2020,rubio2020global,Potts2019fundamentallimits,PhysRevResearch.2.033394}.
	When probes equilibrate with the sample the answer to (i) is given by the heat capacity of the probe,
	and as for (ii), energy measurements are the most precise ones~\cite{PhysRevLett.114.220405}. Nonetheless, we know from the theory of open quantum systems that probes rarely thermalise with the sample that they interact with~\cite{breuer2002theory,weiss2012quantum}. This phenomena is specifically relevant at low temperatures and/or strong probe-sample coupling. 
	Therefore, the impact of non thermalising probes in thermometry, specifically of Bosonic gases, has been considered in several works. For example, it is shown that depending on the temperature regime, strong probe-sample interaction can be beneficial or disadvantageous for thermometry~\cite{PhysRevA.96.062103,PhysRevLett.122.030403}. The extremely determinant role of spectral density at low temperature thermometry was considered in~\cite{PhysRevB.98.045101}, and the impact of bath induced correlations present at low temperatures were addressed in~\cite{planella2020bath}.
	This work is dedicated to thermometry of {\it homogeneous} or uniform ultracold Bosonic gases, in \textit{any spacial dimensionality}. We provide answers to the fundamental questions (i) and (ii) and address the practical problem of (iii) designing experimentally feasible measurements that perform fairly close to optimal. 
		
Homogeneous ultracold gases represent a unique platform to test fundamental quantum phenomena, and have a high potential for quantum technologies. 
They can be realized experimentally in very versatile and extremely controlled set-ups. First realizations were obtained in a uniform three-dimensional optical box trap, formed by a {\it tube} laser beam and two perpendicular {\it sheet} laser beams \cite{2013GauntPRL} (forming a dark optical trap, see \cite{2010JaouadiPRA,2002KaplanJOSAB}).  
This experiment allowed for a thorough characterization of the properties of the gas, such as the condensate interaction energy \cite{2014GotlibovychPRA}, or the critical point for condensation \cite{2014SchmidutzPRL}. A similar technique was used in \cite{2014CormanPRL} to realize uniform Bose gases in a two dimensional box-like potential, which is confined in an  annular geometry contained between an external ring and an inner disk. Uniform ultracold bose gases in a variety of two-dimensional  configurations (e.g.  disc, rectangle, double rectangle), were further explored in \cite{2015ChomazNatcom}. The establishment of this neat experimental tools paved the way to study a large collection of fundamental phenomena in uniform gases e.g., supercurrents~\cite{2014CormanPRL}, the Kibble-Zurek mechanism \cite{2014CormanPRL, 2015ChomazNatcom,2015NavonScience,2017BeugnonJPB}, the power-law scaling of the coherence length \cite{2015NavonScience}, the quantum version of the Joule-Thomson  effect \cite{2014SchmidutzPRL}, several aspects of the depletion and quasi-particle excitations \cite{2017LopesPRLb,2017LopesPRL} or  giant vortex clusters recently \cite{2019GauthierScience}. 

Experiments in uniform Bose-Einstein condensates have thus shown that they are outstandingly controllable systems which allow to test many fundamental effects. On top of this, there is a large number of theoretical proposals which take advantage of their properties for which the uniformity of the gas and the associated universality of the long-wave behavior is crucial. Moreover, the great uniformity of this system makes it a great set-up for correlation studies for very low temperatures ($T\to0$ see \cite{2014GotlibovychPRA}). To date most theoretical works in thermometry of cold gases have fallen short in being applied to experiments, firstly, because the theory involves over-simplifications as they consider toy models as a proof of concept. 
Secondly, due to technological limitations the thermometry schemes with ultimate precision are not experimentally feasible. 
Nonetheless, probe based thermometry of ultracold gases with sub-optimal precision have been experimentally realised in particular cases \cite{PhysRevX.10.011018}.

Our proposal is therefore of large interest in studies of uniform BECs due to its non-destructive feature, rigorous modeling, and high precision with experimentally realisable measurements. Such temperature estimation will boost e.g., characterisation of phase diagrams \cite{2000ReppyPRL, 2001ArnoldPRL, 2001KashurnikovPRL, 2003kleinertMPLB, 2004KasteningPRA, 2005KleinertAnnPhys, 2005YukalovLPL}, or help to account for thermal fluctuations and separate them from the quantum ones. Furthermore, this thermometer keeps its accuracy at very low temperatures, a regime in which the uniform gases promise a large number of interesting effects and applications (e.g. this system has been proposed as a quantum simulator of the early universe \cite{2013OpanchukAnnPhys,2015FialkoEPL,2021NgPRX} or for    relativistic quantum metrology, as acceleration produces observable relativistic effects on homogeneous BECs \cite{2014AhmadiSciRep}).
\begin{figure}
	\includegraphics[width=\linewidth]{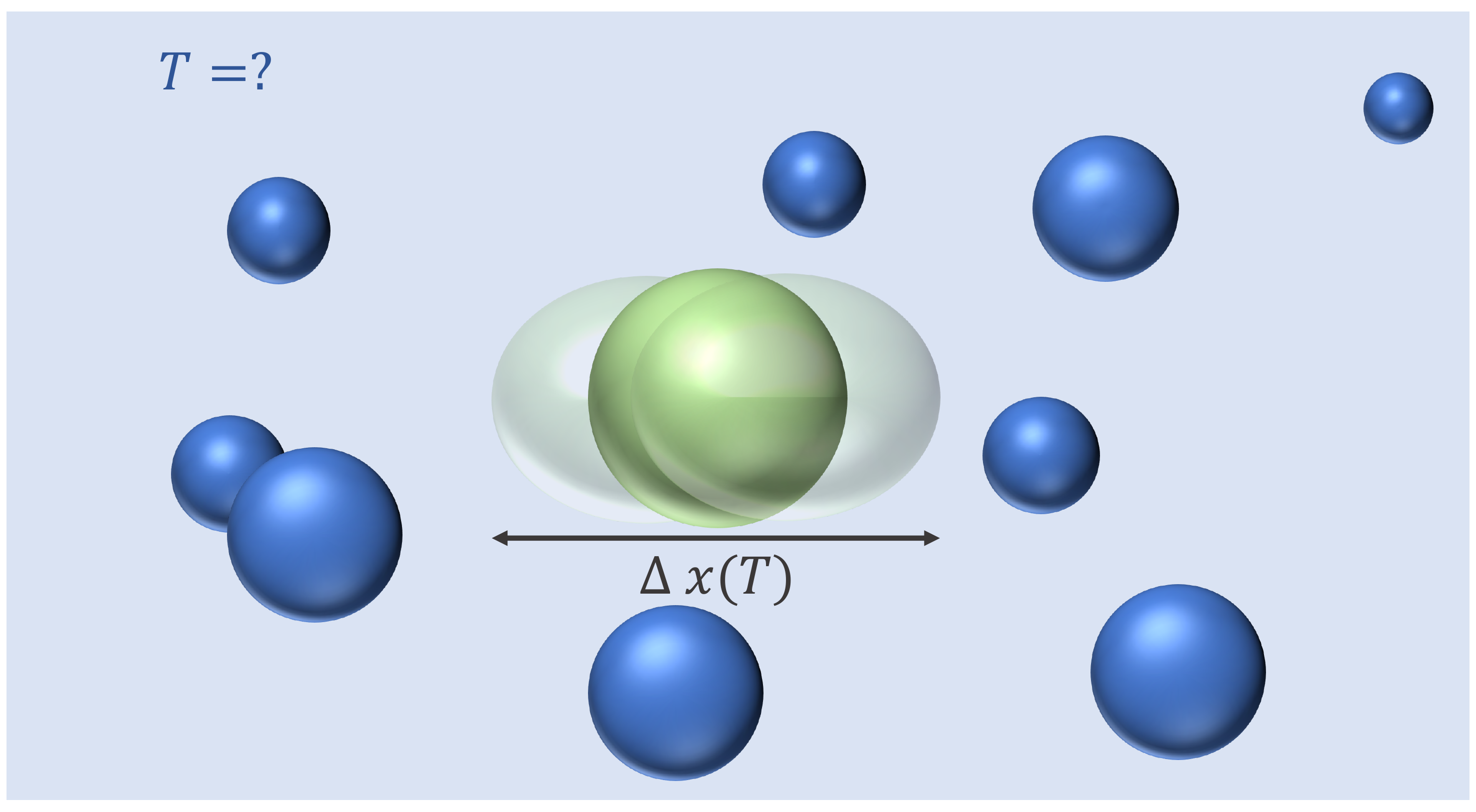}
	\caption{Schematic of our proposed thermometry protocol. The sample is a collection of BEC atoms (represented by the blue balls) trapped in a uniform potential/box with arbitrary dimension, $d\in\{1,2,3\}$. An impurity atom (represented by the green ball) plays the role of our thermometer. The impurity is trapped in a harmonic potential (not shown here). After allowing for sufficiently long interaction, the quantum state of the impurity atom acquires information about the temperature $T$ of the sample. By measuring some property of the impurity e.g., its position fluctuations, one can infer $T$ with minimally disturbing the BEC sample.}
	\label{fig:setup}
\end{figure}	
Particularly, our first-of-its-kind exploration of impurity-based thermometry in one, two and three dimensional homogeneous  Bosonic gases suggest that the ultimate precision substantially improves by increasing the spatial dimension, and thus giving more importance to the thermal fluctuations in $3d$ Bosonic gases. As an example, one can estimate the \textit{sub}-{\rm nK} temperature of a $3d$ Bosonic gas in a homogeneous trap with a relative error below $13\%$ with as few as $100$ measurement runs, whereas in the $1d$ scenario one needs more than $500$ measurements for the same precision. 
We further explore more practical measurements, namely based on the position and momentum of the impurity. Indeed, they can perform good compared to the ultimate bound: For the same target precision they may need twice as many measurements as the optimal (but experimentally complex) measurement. For the $1d$ scenario and at low enough temperatures the position measurement is {\it de facto} optimal. 
	\section{Impurity dynamics in a BEC $-$}\label{sec:LinRes}
	Our setup is illustrated in Fig.~\ref{fig:setup}. We consider an impurity atom with mass $m_{\rm I}$ in an external potential $U_{\rm ext}(\bf r)$. The impurity atom is immersed within a hosting BEC. We consider a homogeneous  BEC of dimension $d$.
	In total, the Hamiltonian describing the non-interacting parts and all the interactions of such a composite system is given by
	\begin{align}
		H= H_{\rm I}+H_{\rm B}+H_{\rm BB}+H_{\rm IB}.
		\label{eq:Hamiltonian}
	\end{align}
	Here, the four terms  represent the non-interacting Hamiltonians of the impurity and of the Bosons, and the Hamiltonians of the Boson-Boson atomic interaction and the impurity-Boson atomic interaction, respectively. Within the second quantization formalism,  their explicit forms are \cite{2017Lampo}
\begin{align}
H_{\rm I}=\frac{\mathbf{P}^2}{2m_{\rm I}}+U_{\rm ext}(\mathbf{r}), \quad H_{\rm B}=\sum_{\mathbf{k}}\epsilon_{\mathbf{k}}a_{\mathbf{k}}^{\dagger}a_{\mathbf{k}},
 \label{eq:HamiltonianI}
\end{align}
 
\begin{align}
H_{\rm BB}=\frac{1}{2V}\sum_{\mathbf{k},\mathbf{k{'}},\mathbf{q}}V_{\rm B}(\mathbf{q})a_{\mathbf{k{'}-q}}^{\dagger}a_{\mathbf{k+q}}^{\dagger}a_{\mathbf{k{'}}}a_{\mathbf{k}},
 \end{align}
\begin{align}
H_{\rm IB}=\frac{1}{V}\sum_{\mathbf{k},\mathbf{q}}V_{\rm IB}\rho_{\rm I}(\mathbf{q})a_{\mathbf{k-q}}^{\dagger}a_{\mathbf{k}}.
 \label{eq:HamiltonianIB}
\end{align}
In $d<3$, the external potential experienced by the Bosons, $V_{\rm ext}(\mathbf{r})$, and that of the impurity, $U_{\rm ext}(\bf r)$, are tightly parabollic in the directions in which the degrees of freedom of the BEC and the  impurity are frozen, so to realize the reduction of dimensionality. In $d=3$, and for $d<3$ in all directions where dimensionality is not reduced, the external potential for the Bosons is zero and  the potential for the impurity $U_{\rm ext}(\bf r)$ is parabolic with trapping frequency $\Omega$.  Then, in these directions, the BEC is homogeneous and contained in spatial domain of  size $V$, say a uniform box in $d=3$ \cite{2015NavonScience}, a quasi-2D uniform potential \cite{2015ChomazNatcom}, or even an annulus geometry \cite{2014CormanPRL}.   
The Bosonic operators $a_{\mathbf{k}}(a_{\mathbf{k}}^{\dagger})$ destroy (create) a Boson of mass $m_{\rm B}$ having wave vector  $\mathbf{k}$ and energy $ \epsilon_{\mathbf{k}}$. The quantities $V_{\rm B}$ and $V_{\rm IB}$ represent the Fourier transforms of the impulsive (contact) Boson-Boson and impurity-Boson interactions respectively. In addition, the impurity density in the momentum space is $\rho_{\rm I}(\mathbf{q})$. The  Hamiltonian can be recast in the form of a Fr\"{o}hlich Hamiltonian, with some constraints over the values of the parameters of the system (like the temperature, trapping frequency $\Omega$ for the impurity, or strength of interactions, see Appendix C in \cite{2017Lampo} and for the trapped case and higher dimensions see \cite{2018LampoPRA,PhysRevA.103.023303}). We call the regime in which  all these conditions are fulfilled the Fr\"{o}hlich regime, and we check in  all calculations presented in this paper that indeed these conditions are fulfilled. Then, in this regime, the Hamiltonian can be written as
\begin{align}
		H=\frac{\mathbf{P}^2}{2m_{\rm I}}+U_{\rm ext}(\mathbf{r})+\sum_{\mathbf{k}\neq 0}E_{\mathbf{k}}b_{\mathbf{k}}^{\dagger}b_{\mathbf{k}}+\sum_{\mathbf{k}\neq 0}\hbar \vect{g}_\mathbf{k}\cdot\mathbf{r}~\mathbf{\pi_{\mathbf{k}}}.
		\label{eq:HamiltonianFrohlichIntFinal}
	\end{align}
In Eq. \eqref{eq:HamiltonianFrohlichIntFinal}, the first term represents the free kinetic energy of the impurity. The last term is the interaction between the position coordinate  $\bf{r}$ of the impurity atom and the Bogoliubov Bosonic modes of BEC. This interaction part is linear when the constraints over the parameters mentioned above are fulfilled \cite{2017Lampo,PhysRevA.103.023303}. This part of the Hamiltonian is written in terms of its canonical dimensionless momenta $\textstyle{\mathbf{\pi_{\mathbf{k}}}=i(b_{\mathbf{k}}-  b_{\mathbf{k}}^{\dagger})}$, with $b_{\mathbf{k}}(b_{\mathbf{k}}^{\dagger} )$ representing the Bogoliubov annihilation (creation) operator for the mode with momentum $ \mathbf{k}$. It thus has the form of the quantum Brownian motion (QBM) model, in which the impurity plays the role of a Brownian particle, while the Bogoliubov modes of BEC act as an effective Bosonic environment. Such an environment forms the multimode thermal states due to the finite temperature of the gas. The coupling constant is given by
	\begin{align}
		\vect{g}_\mathbf{k}=\frac{\mathbf{k}V_{\mathbf{k}}}{\hbar},  \quad  V_{\mathbf{k}}=g_{\rm IB}\sqrt{\frac{n_{0}}{V}}\left[\frac{(\xi k)^{2}}{(\xi k)^{2}+2}\right]^{\frac{1}{4}}.
		\label{eq:CouplingParameterSytemBath}
	\end{align}
	In the above expressions,  the coherence length and the speed of sound are respectively given as
	\begin{align}
		\xi=\frac{\hbar}{\sqrt{2g_{\rm B}m_{\rm B}n_{0}}}~, \quad c=\frac{\hbar}{\sqrt{2}m_{\rm B}\xi}.
		\label{}
	\end{align}
	Within the linear Bogoliubov dispersion relation, it has been shown that the bath is conveniently described  by the super-ohmic spectral density tensor $\underline{\underline{J_{\rm d}}}(\omega)= d^{-1}\pr{\mathcal{J}_{\rm d}(\omega)} \underline{\underline{I} }~_{\rm d\times d}$ for the dimension $d$ of the hosting BEC \cite{PhysRevA.103.023303}. In this expression, the spectral function $\mathcal{J}_{\rm d}$ is shown to take the super-ohmic form
	\begin{align}
		\mathcal{J}_{\rm d}(\omega)=m_{\rm I}\pc{\tau_{\rm d}}^{d}\omega^{d+2}\mathcal{K} (\omega,\Lambda_{\rm d}),
		\label{eq:Spectral DensityFuctionApprox}
	\end{align}
	while we introduce the sharp ultraviolet cut-off $\mathcal{K}$ with a cut-off frequency $ \Lambda_{\rm d}$. This is customary to avoid divergences due to the the rising behaviour of the spectral density at high frequencies.
	Here the $d$-parametrised characteristic time $\tau_{\rm d}$ is given by
	\begin{align}
		\pc{\tau_{\rm d}}^{d}=\frac{S_{\rm d} \pc{\eta_{\rm d}}^{2}}{2(2\pi)^dm_{\rm I}}\left(\frac{m_{\rm B}}{\pr{g_{\rm B, d}}^{\pr{\frac{d}{d+2}}}n_{0,\rm d}}\right)^{(\frac{d+2}{2})}. 
		\label{eq:Taud}
	\end{align}
	For $ d=1, ~2$ and  $3$ we have $S_{1}=2,~ S_{2}=2\pi$ and $S_{3}=4\pi$ respectively, while the $d$-parametrised bath characteristic frequency is $ \Lambda_{\rm d}=(g_{\rm B,d}n_{0,{\rm d}})/\hbar$.  We have additionally written the impurity-Boson coupling in the units of the Boson-Boson coupling as $\eta_{\rm d}=(g_{\rm IB,d}/g_{\rm B,d})$. In addition, the $d$-dimensional density is $n_{0,\rm d}=\pc{n_{0,1}}^d$. The  external potential for the BEC atoms, which contains the transverse harmonic confinement of BEC for $d<3$ that realizes the reduction of dimensionality is explicitly given by 
	\begin{align}
		V_{\rm ext}(\mathbf{r})=\begin{cases}
			(1/2)m_{\rm B}\omega_{\perp}^{2}\left(y^{2}+z^{2}\right) ,\quad  \text{for } d=1\\
			(1/2)m_{\rm B}\omega_{z}^{2}\left(z^{2}\right),\quad  \quad  \text{for }~ d=2\\
			0~ ,\quad \quad\quad \quad \quad \quad \quad  \text{for }~ d=3.\\
		\end{cases}
		\label{}
	\end{align}
The Boson-Boson coupling $g_{\rm B,d}$ is written in terms of the external trap frequency $ \omega_{\rm d}=\px{\omega_{1}=\omega_{\perp},\omega_{2}=\omega_{z}, \omega_{3}=0}$ for the dimension involved, and it has the expression \cite{PhysRevA.70.013608, PhysRevLett.116.225301}
	\begin{align}
		g_{\rm B,{\rm d}}=\frac{S_{\rm d}\hbar^{2}a_{3}}{m_{\rm B}\pc{\sqrt{\hbar/m_{\rm B}\omega_{\rm d}}}^{3-d}}. 
		\label{eq:BosonCouplingConstant&density}
	\end{align}
This makes the cases $d=1$  and $d=2$ to be quasi-$1d$ and quasi-$2d$ respectively. In Ref. \cite{PhysRevA.103.023303},  the above bath characterisation has been employed to derive the equations of motion (EOM) of the impurity position coordinates, while taking into account the bath memory effects. This is accounted by the time-non-local form of the damping kernel $\Gamma_{\rm d}^{xx}$, with $x$ in one of the relevant directions (in $d=2$ and $d=3$ similar expressions are found in the other directions).  The harmonically bond impurity motion   is driven by the effective Brownian stochastic force $B^{x}(t)$, which is further formed by the Bogoliubov modes of the BEC. The corresponding EOM reads
	\begin{align}
		&\ddot x(t)+\Omega^{2} x(t)+\partial_{t}\int_{0}^{t}\Gamma_{\rm d}^{xx}(t-s)x(s)ds\nonumber\\
		&=\pc{\frac{1}{m_{\rm I}}}B^{x}(t),
		\label{eq:EOMPositionImpurityXdirection}
	\end{align} 
	with $x$ representing the position of the impurity in one of the relevant directions. The solution of this equation is shown to take form
	\begin{align}
		x(t)=&G_{\rm 1,d}(t)x(0)+G_{\rm 2,d}(t)\dot x(0)\nonumber\\
		&+(1/m_{\rm I})\int_{0}^{t}dsG_{\rm 2,d}(t-s)B^{x}(s),
		\label{eq:SolutionEOM}
	\end{align}
	where we formulate the Green's functions $G_{\rm 1,d}$ and $G_{\rm 2,d}$ through their Laplace transforms 
	\begin{align}
		&\mathcal{L}_{\mathcal{S},{\rm d}}\left[G_{\rm 1,d}(t)\right]=\frac{\mathcal{S}}{\mathcal{S}^{2}+\Omega^{2}+\mathcal{S}\mathcal{L}_{\mathcal{S}, {\rm d}}\left[\Gamma_{\rm d}^{xx}(t)\right]},\\
		& \mathcal{L}_{\mathcal{S},{\rm d}}\left[G_{\rm 2,d}(t)\right]=\frac{1}{\mathcal{S}^{2}+\Omega^{2}+\mathcal{S}\mathcal{L}_{\mathcal{S}, {\rm d}}\left[\Gamma_{\rm d}^{xx}(t)\right]}.
		\label{eq:LaplaceG1}
	\end{align}
	By introducing the sharp cut-off given by $ \mathcal{K}=\Theta (\Lambda_{\rm d}-\omega)$, with $\Theta$ is the Heaviside step function, the $d$-parametrised damping kernel reads
	\begin{align}
		\mathcal{L}_{\mathcal{S},{\rm d}}\left[\Gamma_{\rm d}^{xx}(t)\right]=\frac{\pc{\Lambda_{\rm d}} ^{d+2} \pc{\tau_{\rm d}} ^d \,_2F_1\left(1,\frac{d+2}{2};\frac{d+4}{2};-\frac{\pc{\Lambda_{\rm d}} ^2}{\mathcal{S}^2}\right)}{d(d+2)\mathcal{S}},
		\label{eq:LaplaceDamping}
	\end{align}
	with $ \,_2F_1\left[.\right]$ denoting the hypergeometric function. 
	
	In the following we aim to use the impurity motion statistical properties to estimate the temperature of the BEC, generalizing to any dimension and homogeneous gases the proposal introduced in  \cite{PhysRevLett.122.030403} for one-dimensional and trapped gases.
	The quadratures of interest are the position and momentum of the impurity. Additionally, the Gaussian nature of its dynamics is fully characterised  by its covariance matrix. In order to quantify the dynamics of the impurity, we start by writing a formal representation of the steady state covariance matrix of the impurity. It is important to note that the steady state covariance matrix is diagonal taking the form $ \bm{\sigma}\equiv{\rm diag}\pc{\Delta x^2, \Delta p^2}$.  Here $\Delta x^2$ and $\Delta p^2$ are the steady state position and momentum variances of the impurity mechanical mode, which are appropriately scaled such that the  uncertainty relation reads $\Delta x \Delta p\geq1$ \cite{PhysRevLett.89.137903}. 
	The fundamental expressions of the position and momentum variances in the unscaled coordinates has explicit connectivity to the imaginary part of the susceptibility, $\tilde{\chi}_{\rm d}^{''}(\omega)$ (fluctuation-dissipation theorem), that is, 
	\begin{align}
		\av{x^2}_{\rm d}=\frac{\hbar}{\pi}\int_{0}^{\infty}d\omega\coth\pc{\frac{\hbar\omega}{2k_{\rm B}T}}\tilde{\chi}_{\rm d}^{''}(\omega),
		\label{eq:SteadyStatePositionVariene}
	\end{align}
	
	\begin{align}
		\av{p^2}_{\rm d}=\frac{\hbar m_{\rm I}^{2}}{\pi}\int_{0}^{\infty}d\omega~\omega^{2}\coth\pc{\frac{\hbar\omega}{2k_{\rm B}T}}\tilde{\chi}_{\rm d}^{''}(\omega),
		\label{eq:SteadyStateMomentumVariene}
	\end{align}
	where 
	\begin{align}
		\tilde{\chi}_{\rm d}^{''}(\omega)=\frac{1}{m_{\rm I}}\frac{\xi_{\rm d}^{xx}(\omega) \omega}{\pr{\omega\xi_{\rm d}^{xx}(\omega)}^{2}+\pr{\Omega^{2}-\omega^{2}+\omega \theta_{\rm d}^{xx}\pc{\omega}}^{2}}.
	\end{align}
	We point out that any kind of {\it{super-Ohmcity}} (and also the dimensional information) is captured here by the functions $\xi_{\rm d}^{xx}(\omega)$  and $\theta_{\rm d}^{xx}\pc{\omega}$ which are respectively the Fourier domain real and imaginary part of the damping kernel (cf. Eq. \eqref{eq:LaplaceDamping}).

	\section{Thermometry of BEC in different dimensions $-$}\label{thermometry}
	Estimation theory, in part, deals with inference of a parameter $\mathcal{P}$ from a set of measurement outcomes with the aim of minimising the estimation error. This set is collected through performing a positive operator-valued measurement (POVM) on the quantum system, with a density matrix $\hat{\rho}\pc{\mathcal{P}}$ that depends on the parameter.
	Let $\{\hat{\Pi}^k_{\mathbf{x}}\}$ represent the POVM elements of the measurement. Here, $k$ denotes the choice of measurement while ${\mathbf{x}}$ labels different outcomes. Then $\hat{\Pi}^k_{\mathbf{x}}\geq 0$ and we have $\textstyle{\int d{\mathbf{x}}~\hat{\Pi}^k_{\mathbf{x}}=\mathbb{I}}$, $\forall k$. 
	The uncertainty (randomness) in the POVM outcomes allows only to infer the parameter with limited precision. According to the Cr\'amer-Rao bound for an unbiased estimator the estimation error---as quantified by the mean squared error and denoted by ${\rm Var}\pc{\mathcal{P},k}$---is lower bounded by the inverse of the Fisher information, that is $\textstyle{{\rm Var}\pc{\mathcal{P},k}\geq\pr{N F\pc{\mathcal{P},k}}^{-1} }$. Here, $N$ is the number of (independent) measurement runs and appears as a result of the central limit theorem. The Fisher information $F\pc{\mathcal{P},k}$ is given by
	\begin{align}
		F\pc{\mathcal{P},k}=\int \pr{ \partial_{\mathcal{P}}~ {\rm ln} ~p({\mathbf{x}}|\mathcal{P},k)}^{2}p({\mathbf{x}}|\mathcal{P},k) d{\mathbf{x}},
		\label{eq:FishInfo}
	\end{align}
with $\textstyle{p({\mathbf{x}}|\mathcal{P},k)={\rm Tr}[ \hat{\rho}\pc{\mathcal{P}}\hat{\Pi}^k_{\mathbf{x}}]}$ being the distribution of the measurement outcomes, conditioned on the true value of the parameter to be $\mathcal{P}$ and the measurement $k$ being performed. Note that the Cram\'er-Rao bound can be saturated using the maximum likelihood estimator. 
	
The Fisher information clearly depends on the specific measurement performed, hence the argument $k$. One may optimise it over all possible measurements to get the so called quantum Fisher information (QFI) that quantifies yet another fundamental lower bound on the estimation error regardless of the measurement. The QFI can be written as \cite{PhysRevLett.72.3439, doi:10.1142/S0219749909004839} 
	\begin{align}
		F^{\rm Q}\pc{\mathcal{P}} \coloneqq \max_{k} F\pc{\mathcal{P},k} = {\rm Tr}\pr{ \hat{\rho}\pc{\mathcal{P}}\hat{\Lambda}^{2}\pc{\mathcal{P}}},
		\label{eq:FishInfo-Sym-Log}
	\end{align}
	where the Hermitian operator $\hat{\Lambda}\pc{\mathcal{P}}$ is the symmetric logarithmic derivative (SLD) defined as
	\begin{align}
		\partial_{\mathcal{P}}\hat{\rho}\pc{\mathcal{P}}\equiv\px{ \hat{\Lambda}\pc{\mathcal{P}},\hat{\rho}\pc{\mathcal{P}}}/2.
		\label{eq:SLD}
	\end{align}
	In fact, a measurement performed in the basis of SLD is optimal.
	
Generally speaking, finding the optimal measurement and/or the ultimate error bound is a challenging task. Nonetheless, as we proved in the previous section, here we are dealing with a Bosonic Gaussian system, which is fully characterised by its displacement vector and covariance matrix. It has been shown that the QFI of Gaussian systems can be written in terms of the covariance matrix  $\bm{\sigma}$, as follows \cite{monras2013phase,PhysRevA.89.032128, PhysRevA.88.040102,  Genoni_2016}
	\begin{figure}[!t]
\includegraphics[width=1\linewidth]{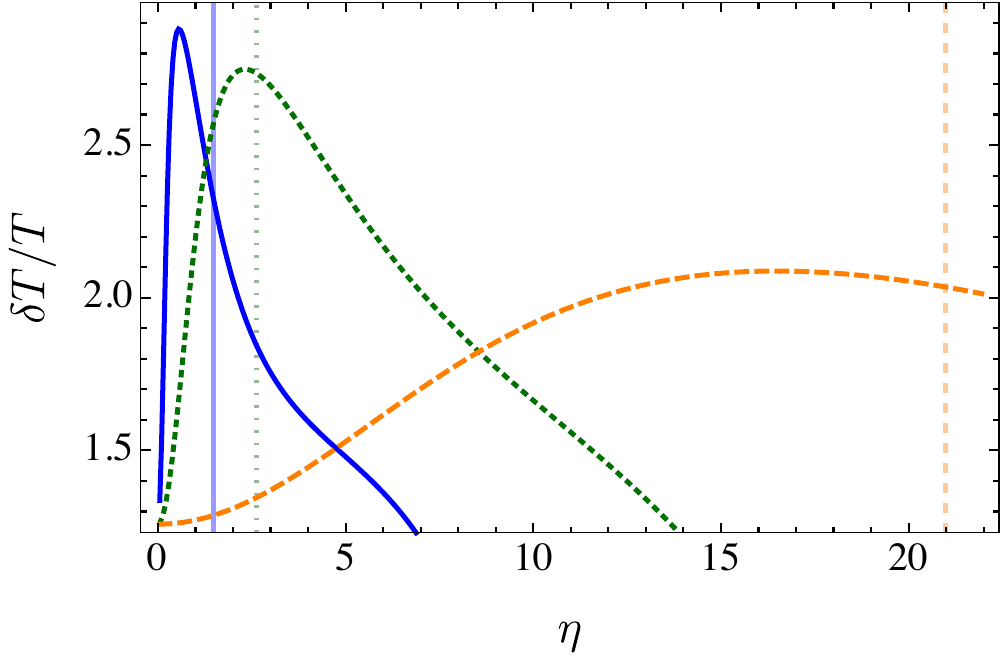} 
\caption{Minimum relative temperature error---obtained by measuring $\hat{\Lambda}$---as a function of the dimensionless system-bath coupling. Solid blue, dotted green, and  dashed orange curves represent the $1d$, $2d$, and $3d$ cases, respectively. The corresponding background vertical lines set the upper bound of the coupling $\eta_{\rm c,d}$ for the validity of the Fr\"{o}hlich regime.  The three dimensional scattering length is $a_{3}=100 a_{0}$. The $1d$ density is kept to be $n_{0,1}=3(\mu{ \rm m})^{-1}$. For the cases of quasi- 1d and 2d settings, we have set Bosons transverse confinement frequency to be  $ \omega_{\perp}=\omega_{z}=2\pi \times 34~ {\rm k Hz}$.  These results correspond to  $T=0.2~{\rm nK}$ for all the dimensions.  The impurity trapping frequency is $\Omega=2\pi \times  10~ \rm Hz$. The number of measurements are $N=1$. These results refers to a potassium ${\rm K}$ impurity atom with mass $m_{\rm I} = 6.4924249\times10^{-26} \rm{kg}$, immersed in a gas of rubidium $\rm Rb$  atoms with  mass $m_{\rm B} = 1.44316060\times10^{-25} \rm{kg}$.}  \label{Therfig:fig1}
\end{figure}
	\begin{align}
		F^{\rm Q}\pc{\mathcal{P}}=\frac{1}{2}\frac{\rm Tr\pr{\pc{\bm{\sigma}^{-1}\pr{\partial_{\mathcal{P}}\bm{\sigma}}}^{2}}}{1+{\mu}^{2}}+2\frac{\pc{\partial_{\mathcal{P}}{\mu}}^{2}}{1-{\mu}^{4}}.
		\label{eq:QFI-Cov1}
	\end{align}
	Here $\mu=1/\sqrt{\Delta x^2 \Delta p^2} $ is the purity function of the impurity state. We particularise this to our thermometry problem with $\mathcal{P}=T$, and use Eqs.~\eqref{eq:SteadyStatePositionVariene} and \eqref{eq:SteadyStateMomentumVariene}. Note that the displacement vector as well as $\av{\left\{x,p\right\}}$ vanish in this case. 
	Let our figure of merit be $\textstyle{ \delta T(k)/T} \coloneqq 1/\sqrt{N F\pc{\mathcal{P},k}}$ which is the relative temperature error using the measurement $k$ and maximum likelihood estimator; the smaller it is, the more precise the thermometer will be. One has $ \textstyle{ \delta T(k)/T} \geq 1/\sqrt{N F^{\rm Q}\pc{\mathcal{P}}} = \delta T (\hat \Lambda)/T\eqqcolon \delta T_{\min}/T $. 
	In what follows we examine the optimal and experimentally feasible sub-optimal measurements in different spacial dimensions.
	\begin{figure}[!t]
		\includegraphics[width=1\columnwidth]{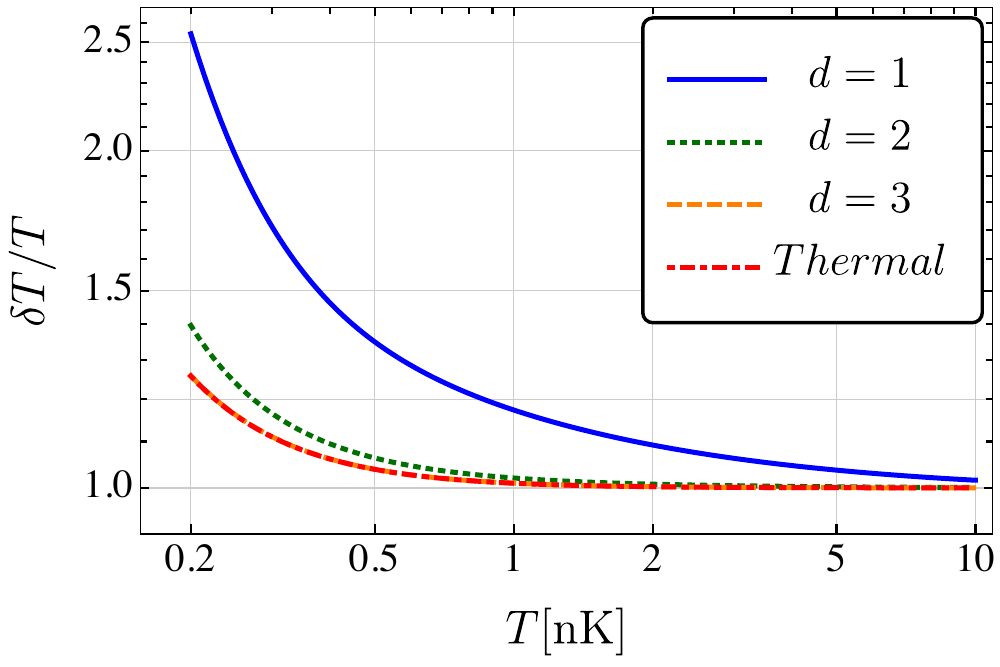}
		\caption{Minimum relative temperature error as a function of the temperature. We set the coupling to $ \eta=0.3$. For comparison, we also plot the error of a thermal state, which matches the $d=3$ case for the range of temperatures considered here. The rest of the parameters are same as in Fig. \ref{Therfig:fig1}.}
		\label{Therfig:fig2}
	\end{figure}
	\subsection{The optimal thermometry and the fundamental bounds}
	In Fig. \ref{Therfig:fig1} we fix the temperature, and show the relative temperature error, $\delta T_{\min}/T $, versus the dimensionless system-bath coupling, $\eta$. The relative error peaks at some coupling $\eta_{\rm p, d}$ that depends on the temperature and the dimensionality. 
	For the regime where $\eta<\eta_{\rm p, d}$, the error increases as we strengthen the system bath coupling. On the other hand, for $\eta>\eta_{\rm p, d}$, the opposite is observed. 
	For the latter case, however, the enhancement of the coupling is limited by the maximum allowed system bath coupling $\eta_{\rm c,d}$
in order to keep the calculations within the Fr\"{o}hlich regime (see caption of the Fig.~\ref{Therfig:fig1}). As shown in Appendix \ref{App:1}, for all the dimensions, the thermal statistics of the impurity is mostly contributed by the ground state for $\eta<\eta_{\rm p, d}$.  Such regime would then reflect a state with high purity. For the case of  $\eta>\eta_{\rm p, d}$ higher order Fock states significantly contribute in the construction of such statistics and the state becomes more mixed. This leads to the results that smaller values of the coupling -- as characterised by $\eta<\eta_{\rm p, d}$ -- are suitable for thermometry only as long as system closely follows the ground state with high purity. Yet, for a strongly coupled bath with the system ($\eta>\eta_{\rm p, d}$) -- where bath induced thermalisation occurs -- the large coupling values give better thermometric performance. However the latter regime is always bounded above by some critical coupling in the present physical setup. Finally from Fig.~\ref{Therfig:fig1}, the peak error is relatively reduced as we move to higher dimensions reflecting that higher dimensions are better for thermometry. 
	\begin{figure*}[!t]
		\includegraphics[width=0.65\columnwidth]{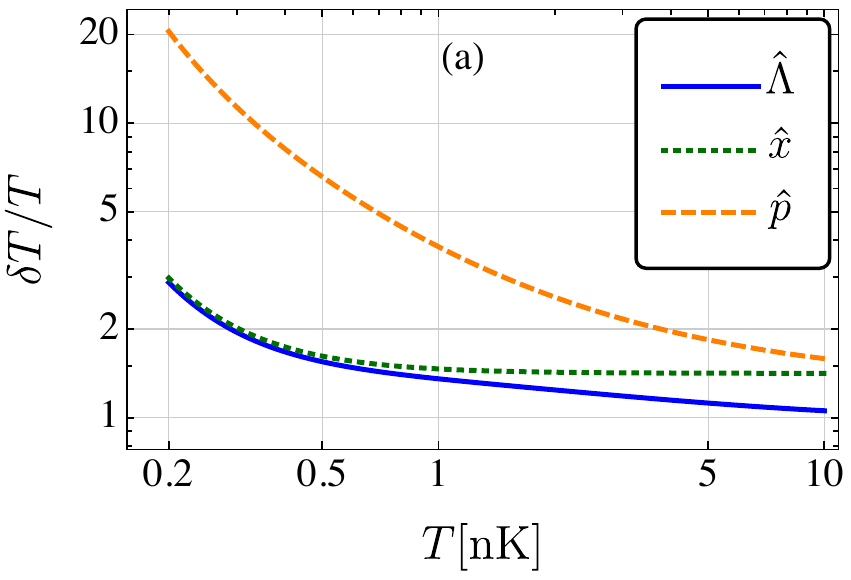}
		\includegraphics[width=0.65\columnwidth]{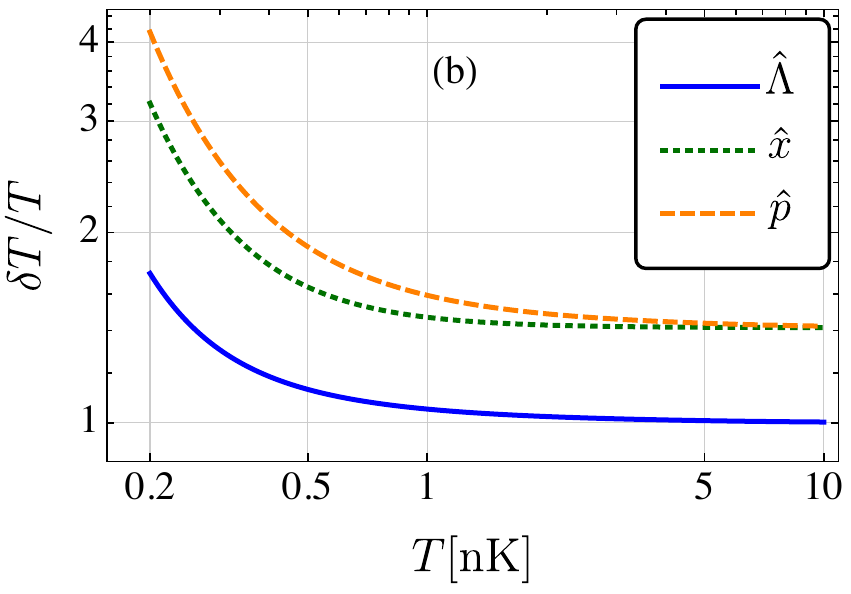}
		\includegraphics[width=0.68\columnwidth]{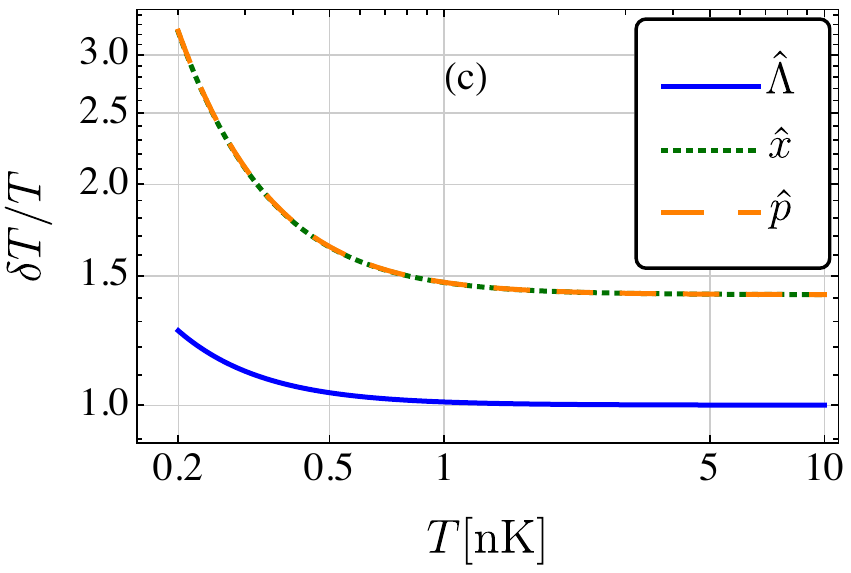}\\
		
		\includegraphics[width=\linewidth]{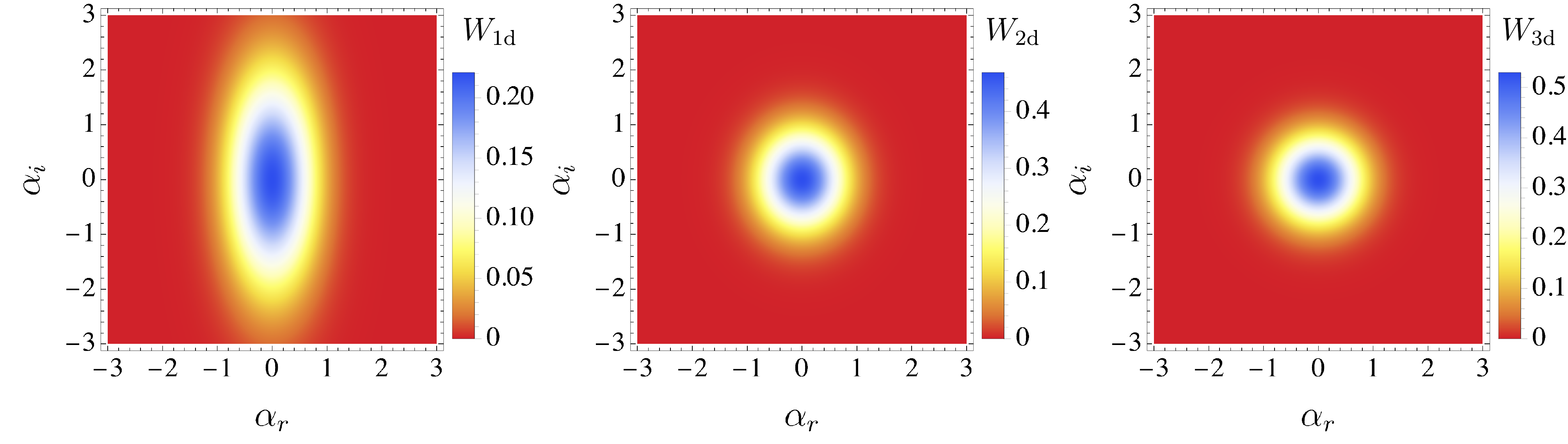} 
		\caption{ Upper panels: Relative error as a function of coupling based on quadrature measurements (${\hat x}$, and ${\hat p}$) as well as the optimal measurement $\hat{\Lambda}$. (a), (b) and (c) represent the  $1d$, $2d$ and  $3d$ cases, respectively. For all plots we set $\eta=0.6$ and $N=1$. Lower panels: corresponding to each case above, the Wigner function of the impurity state is plotted for $ T= 0.2 {\rm nK} $. In the $1d$ case, an enhanced position squeezing is obtained, while as we move to higher dimensions, we obtain lesser position squeezing effect.}
		\label{Therfig:fig4}
	\end{figure*}
In order to further analyse the performance of the thermometer, we plot the relative temperature error versus the true temperature of the hosting BEC for different dimensions in Fig.~\ref{Therfig:fig2}. 
For temperature range of interest---$0.2{\rm nK}\leq T \leq 10{\rm nK}$---higher $d$ always leads to smaller error. This is an evidence that for high enough temperatures larger non-ohmicity (which is related to higher dimension) leads to better thermometry precision. In \cite{PhysRevA.96.062103}, a similar behaviour is reported. There, it is also shown that at extremely low temperatures, the opposite behaviour is observed i.e., lower non-ohmicity leads to higher precision. Nonetheless, for our BEC setup such extreme lows are irrelevant as they fall even below the pico Kelvin regime. 
For $d=3$ scenario, we observe that the $\delta T_{\min}/T\simeq1.25$ at $T = 0.2{\rm nK}$ for a single shot. Thus with $N=100$ measurements one gets a relative error of $\simeq 12.5 \%$. In $d=1$ the same precision requires about $N=400$ measurements. For temperatures $T\gtrsim 1{\rm nK}$ the relative error in two and three dimensions saturate to $\delta T_{\min}/T\simeq 1$ i.e., one can have $10\%$ error with $N=100$ measurements. This is in fact a result that one expects when the impurity is thermalised and the ratio $T/\Omega \gg 1$. 
	 For the one dimensional scenario, the saturation of error only appears for $T\gtrsim 10 {\rm nK}$. We also show the relative error down to $50 {\rm pK}$ in appendix~\ref{App:2}, serving as a guide to the eye about the error enhancement at low temperature.

	\subsection{Sub-Optimal thermometry and practical measurements}\label{MarkMeasure}
	Now that we have found the ultimate precision bounds on estimating the temperature of the BEC, we move forward to examine the estimation error by measuring more practical observables, namely an individual measurements of either on position or momentum of the impurity~\cite{PhysRevLett.107.020405}. 
	
We note that both position and momentum belong to the family of Gaussian measurements, and can be fully characterised by an associated covariance matrix $\bm{\sigma}^{k}$, with $k\in\{{\hat x},{\hat p}\}$ labeling the measurement. In particular, we have $\bm{\sigma}^{\hat x} = \lim_{R\to\infty} {\rm diag}(1/R,R)$, and $\bm{\sigma}^{\hat p} = \lim_{R\to\infty} {\rm diag}(R,1/R)$, with $R$ being some squeezing parameter. When performed on a system with covariance matrix $\bm{\sigma}$, the corresponding Fisher information of these measurements is given by \cite{Cenni_in_preparation}
	\begin{align}
		F(T,k) = \frac{1}{2}{\rm Tr}\left[ \left( (\bm{\sigma}^{k} + \bm{\sigma})^{-1}\partial_T (\bm{\sigma}^{k} + \bm{\sigma})\right)^2 \right].
	\end{align}
	By doing some  algebra we find
	\begin{align}
		F(T,{\hat x}) & = \frac{|\partial_T \av{{\hat x}^2}|^2}{2\av{{\hat x}^2}^2} = \frac{|\partial_T \av{{\hat x}^2}|^2}{{\rm Var}({\hat x}^2)},
	\end{align}
	which is nothing but the (inverse) of error propagation for the observable ${\hat x}^2$. A similar relation connecting the Fisher information of measuring ${\hat p}$ and the error propagation of the observable ${\hat p}^2$ holds by changing ${\hat x}\to {\hat p}$. 
	
In Fig. \ref{Therfig:fig4} (a-c) we depict, for all dimensions, the relative temperature error by measuring position and momentum and compare them to the optimal -- yet challenging -- measurement. Quite notably, in the quasi-$1d$ setting, the position based measurement reaches the optimal profile in the low temperature regime i.e. below $T\lesssim 1 {\rm nK}$. 
This is because of the notable position squeezed states of the impurity motion in the $1d$ case. As such, the Fisher information scales as $F_{\rm Q, x}\pc{T}\sim 1/\Delta x^4$. Therefore the small amount of noise in the position would lead to an enhanced value of the Fisher information and hence a smaller value of the relative error. However, this is obtained in the expense of the error enhancement in the momentum based measurement. The squeezing effects are shown through the plotted Wigner function (for its expression, see Appendix \ref{App:1}) of the impurity state in Fig. \ref{Therfig:fig4}.  For higher dimensions, there is less position squeezing effect and either of the measurements would result in the equivalent amount of error.  For the $2d$ case, the separation of the error between position and momentum becomes lesser and both of them are always away from optimality, but within the same order of magnitude as the optimal measurement. For the $3d$ case the relative error for position and momentum measurements is equal, which is a reflection of the fact that both quadratures have the same temperature dependence. In this dimension, the sub-optimal measurements are not quite effective in extreme lows, but for $T\gtrsim 1 {\rm nK}$ they are within the same order of magnitude as the optimal measurement. Lastly, let us comment that as a result of thermalisation at high enough temperatures ($T\gtrsim 10 {\rm nK}$ for $1d$, $T\gtrsim 5 {\rm nK}$ for $2d$, and $T\gtrsim 1 {\rm nK}$ for $3d$) one has $\delta T(k)/T \approx \sqrt{2} \delta T_{\min}/T=\sqrt{2} $ for $k\in\{\hat{x},\hat{p}\}$. This means with measuring position/momentum twice as many times as the optimal measurement one can achieve the same target precision.

Perhaps the easiest way of performing the measurement of  mean square displacement or momentum fluctuations could be done in the following scenario:  
Instead of looking at a single impurity, we can consider, say, $1000$ to few thousands of non-interacting impurities being polarized cold Fermionic atoms. This idea contradicts in a sense the principal idea of the sample-probe interactions, where the probe size is significantly smaller than the sample, and that the sample is not directly measured. Still, with a condensate, say, $10^7$ Bosons, few thousands probing, non-interacting Fermions will not make much difference. On the other hand, both their spatial and momentum density distributions are easily accessible experimentally: the former either through non-destructive light (refractive index) imaging or, more sophisticated, atomic microscopy, the latter through opening the trap for Fermions and looking at the momentum distribution in the time-of-flight measurement \cite{Lye_1999, Andrews84}.   
\section{Conclusion $-$}\label{MarkMeasure}
Using an individual impurity as a temperature probe is crucial in non-demolition thermometry of cold gases; it has been studied theoretically in abstract models as well as more comprehensive ones in both Fermionic and Bosonic gases~\cite{PhysRevLett.122.030403,PhysRevLett.125.080402,onofrio2016physics,planella2020bath,PhysRevA.96.062103}. Recently, the idea was even put into experimental examination in the mili Kelvin thermometry of Fermionic gases \cite{PhysRevX.10.011018}. 
	The results of the current work pave the path towards implementation of impurity based thermometry in more versatile experimental setups of Bosonic gases in different spacial dimensions $d=1$, $d=2$, and $d=3$, for example \cite{Catani2012,Spethmann2012,2015Hohmann,2018Schmidt,Rentrop2016,2018Schmidt,2013Scelle,2010Zipkes,Jorgensen2016,Hu2016,Yan2019}.  
	
	Our results show that, for the temperature range $0.2 {\rm nK} \leq T \leq 10 {\rm nK}$, as one decreases the spatial dimension $d$ the relative error reduces. Specifically, in the $3d$ scenario and for temperatures as small as $T\approx 0.2 {\rm nK}$ one can achieve the minimum relative error of $12.5\%$ by $N=100$ measurements. To achieve the same precision in the $1d$ case, one should perform more than $N=400$ measurements. It must be noted that these are the fundamental bounds characterised by the quantum Cram\'er-Rao bound, they are in principle achievable i.e., there exist physical measurements that can achieve such precisions. Nonetheless, experimental limitations will not allow for their realisation. What is more doable, are sub-optimal Gaussian measurements, namely measuring position and momentum of the impurity. 
	
	In the $3d$ case and for $T\leq 0.5 {\rm nK}$ these measurements should be repeated an order of magnitude more than the optimal measurement in order to reach the same precision. For the $2d$ scenario one only requires few times more repetitions, while for the $1d$ scenario the position measurement is optimal. At higher temperatures, both position and momentum measurements perform very good, and their relative error can be as good as the minimum error with twice as many measurements. 
	
	Apart from the experimental realisation of our analysis, a possible future direction is the theoretical developement of thermometry of Bosonic gases by means of dynamical probes, i.e., measuring the probe before it reaches the steady state. This can be of more use in scenarios with long thermalisation time. Another interesting problem is to take the temperature of pure Bosonic gases, aligned with the recently proposed framework of~\cite{mitchison2021taking}.

	\section*{Acknowledgments}
	M.M.K. and H.T. acknowledge support from Funda\c{c}\~ao para a Ci\^encia e a Tecnologia (FCT-Portugal) through Grant No PD/BD/114345/2016 and through Contract No IF/00433/2015 respectively. M.A.G.M. acknowledges funding from the Spanish Ministry of Education and Vocational Training (MEFP) through the Beatriz Galindo program 2018 (BEAGAL18/00203).  We (M.L. group) acknowledge the Spanish Ministry MINECO (National Plan 15 Grant: FISICATEAMO No. FIS2016-79508-P, SEVERO OCHOA No. SEV-2015-0522, FPI), European Social Fund, Fundació Cellex, Fundació Mir-Puig, Generalitat de Catalunya (AGAUR Grant No. 2017 SGR 1341, CERCA program, QuantumCAT\_U16-011424, co-funded by ERDF Operational Program of Catalonia 2014-2020), MINECO-EU QUANTERA MAQS (funded by The State Research Agency (AEI) PCI2019-111828-2 / 10.13039/501100011033), and the National Science Centre, Poland-Symfonia Grant No. 2016/20/W/ST4/00314. M.M. acknowleges financial support from the Swiss National Science Foundation (NCCR SwissMAP)

\bibliography{References1.bib}
	
	\appendix
	\section{Thermal statistic of the impurity  $-$} \label{App:1}
	In order to calculate the thermal statistics of the impurity we formally write the density operator of the impurity $\hat{\rho}_{\rm I}$. Any physical state being represented by the density operator is bounded since its Hilbert Schmidt norm is finite. It is then possible to expand a Boson mode state in the basis $\lbrace\hat{D}^{\dagger}(\lambda)\forall\lambda\in\mathbb{C}, \lambda=\lambda_{r}+i\lambda_{i} \rbrace$. Hence one can write the state of the impurity as \cite{PhysRev.177.1882}
	\begin{equation}\label{singledensity}
		\hat{\rho}_{\rm I}= \frac{1}{\pi^2}\int d^{2}\lambda ~ \chi(\lambda,\lambda^{\ast})\hat{D}^{\dagger}(\lambda).
	\end{equation}
	Here $\hat{D}^{\dagger}(\lambda)$ is the conjugate transpose of the coherent displacement operator $\hat{D}(\lambda)=e^{\lambda \hat{a}^{\dagger}-\lambda^{\ast}\hat{a}}$ and $d^{2}\lambda=d\lambda_{r} d\lambda_{i}$ is the measure for two-dimensional integral corresponding to real part $\lambda_{r}$ and imaginary part $\lambda_{i}$ of the complex variable $\lambda$. For a canonical position $\hat{x}$ and momentum $\hat{p}$ variables associated to the harmonic oscillator, they can be expressed in terms of the quadrature $\textstyle{\hat{x}=\pc{\hat{a}+\hat{a}^\dagger}/\sqrt{2}}$ and~$\textstyle{\hat{p}=i\pc{\hat{a}^\dagger-\hat{a}}/\sqrt{2}}$ respectively. The commutation relation therefore reads  $\pr{\hat{x},\hat{p}}=i$.  Here  $\hat{a}(\hat{a}^\dagger)$ is the annihilation (creation) operator for the Bosonic mode. In such coordinates $\textstyle{ \alpha=\pc{1/\sqrt{2}}(\av{\hat{x}}+i \av{\hat{p}})}$. This reflects that the $\alpha_{r}$ and $\alpha_{i}$ are simply the scaled position and momentum expectation value and therefore a valid representation of the phase space. The symmetric ordered  characteristic function $\chi(\beta,\beta^{\ast})$ is the expectation value of the Weyl operator, given by $\textstyle{\chi(\beta,\beta^{\ast})={ \rm Tr}[\hat{\rho}_{\rm I} \hat{D}(\beta)]}$. If the dynamics is Gaussian, the quantum characteristic function is fully captured by the covariance matrix by the expression $\chi\pc{\beta,\beta^{\ast}}= e^{-\pc{1/2}\textbf{x}~\bm{\sigma}~\textbf{x}^{T}}$. Here $\bm{\sigma}$ is the covariance matrix associated to the mode and  $\textbf{x}=\pc{\beta_{i}, \beta_{r}}$ are the phase space variables \cite{PhysRevLett.84.2722}. The Wigner quasi probability distribution  function  $W(\alpha, \alpha^{\ast})$ is the two dimensional complex Fourier transform of the characteristic function such that
	\begin{figure*}[!t]
		\includegraphics[width=0.325\linewidth]{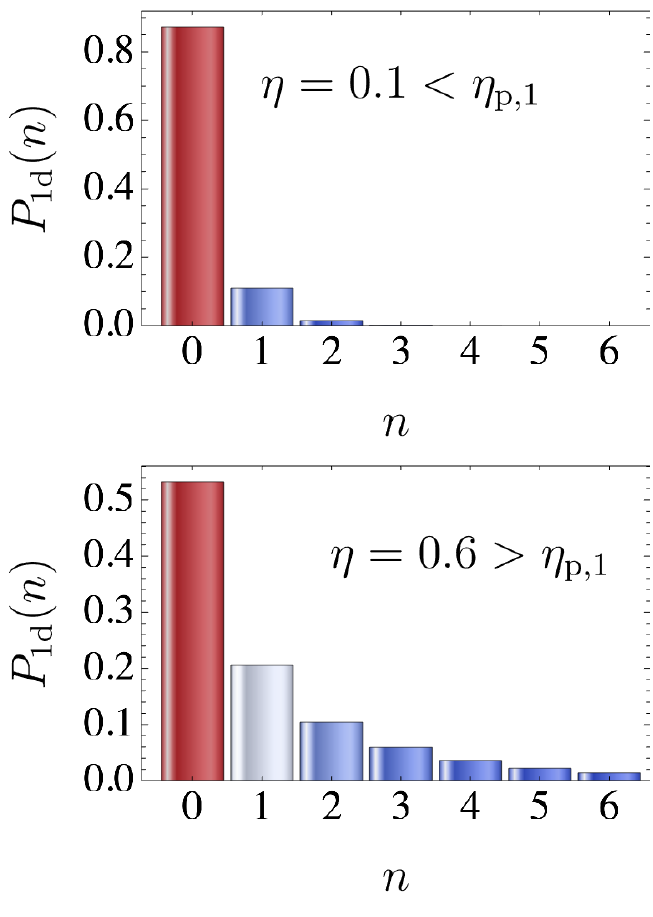} 
		\includegraphics[width=0.325\linewidth]{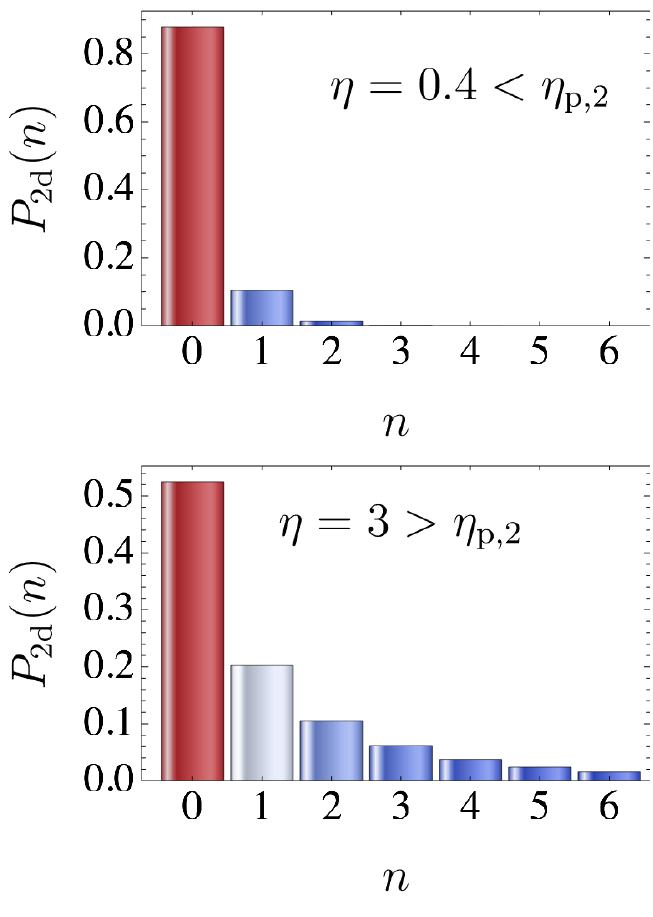}
		\includegraphics[width=0.325\linewidth]{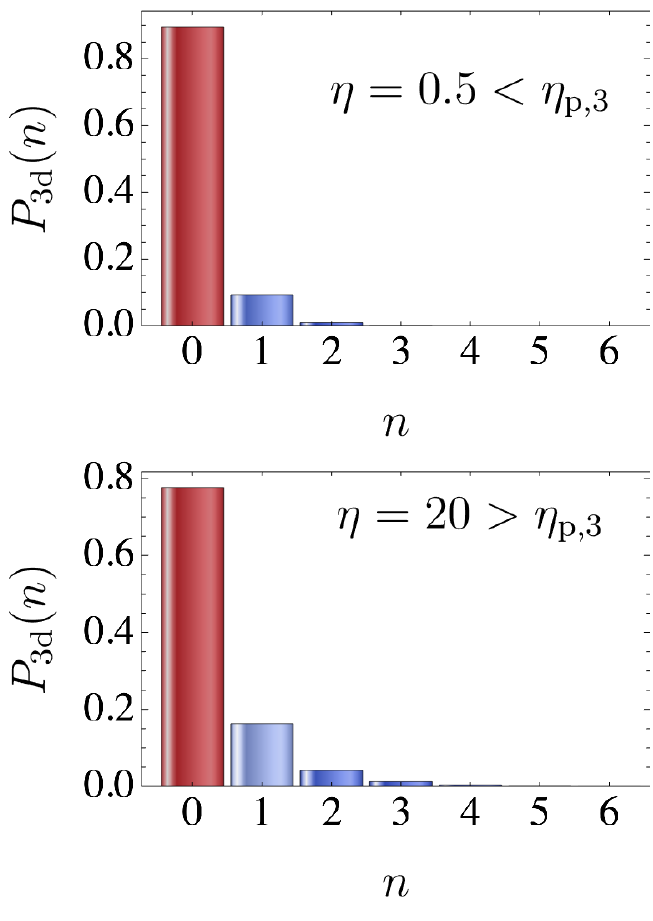}
		\caption{Phonon number distribution of the impurity motion. Left, middle and right columns represent the $1d$, $2d$ and $3d$ cases respectively. The chosen couplings regimes and their values are shown in the inset of each of the figure. Rest of the parameters are the same as in Fig. \ref{Therfig:fig1}.}  \label{Therfig:fig3}
	\end{figure*}
	\begin{align}
		W(\alpha, \alpha^{\ast})=\frac{1}{\pi^{2}}\int d^{2}\beta \chi(\beta,\beta^{\ast}) e^{-\pc{\beta\alpha^{\ast}-\beta^{\ast}\alpha}}.
	\end{align} 
	In addition,  the phonon number statistics, that is, the probability of finding $n$ phonon in the mode, is connected to the density operator by the expression
	$P(n)=\langle n|\hat{\rho}_{\rm I}|n\rangle$. This can be calculated via the matrix elements of the displacement operator being evaluated in the Fock basis and given by the expression \cite{PhysRevA.41.2645}
	\begin{align}
		\langle n|\hat{D}(\alpha)|m\rangle=\sqrt{\frac{m!}{n!}}e^{-|\alpha|^2/2}(\alpha)^{n-m}\mathcal{L}_{m}^{n-m}\pr{|\alpha|^2}, ~n\geq m,
	\end{align}
	where $\mathcal{L}_{m}^{n-m}[.]$ are the generalized Laguerre polynomials. In the present case, we are mainly interested in the diagonal entries therefore we have $m=n$. 
	We plot the impurity phonon number distribution both for $\eta<\eta_{\rm p,d}$ and $\eta>\eta_{\rm p,d}$ for each  dimension in Fig. \ref{Therfig:fig3}. For the $\eta<\eta_{\rm p,d}$ thermal statistics of the impurity are almost fully contributed by the ground state while for $\eta>\eta_{\rm p,d}$, higher order Fock states contribute significantly in the construction of the thermal phonon statistics. 
		%
\section{Relative error for extended temperature range down to {\rm 50 pK}$-$} \label{App:2} 
\begin{figure}[!h]
\includegraphics[width=1\linewidth]{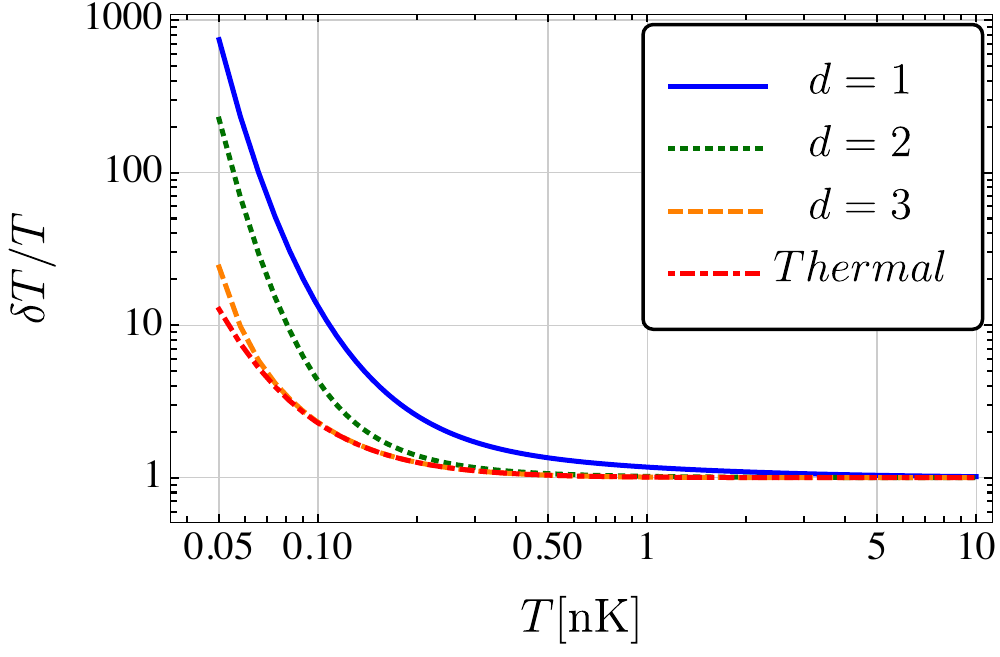} 
\caption{Same as in Fig. \ref{Therfig:fig2}, only the temperature domain has been extended within the interval $0.05{\rm nK}\leq  T\leq 10{\rm nK}$.}  \label{AppB:Supfiga}
\end{figure}
	\begin{figure*}[!h]
		\includegraphics[width=0.68\columnwidth]{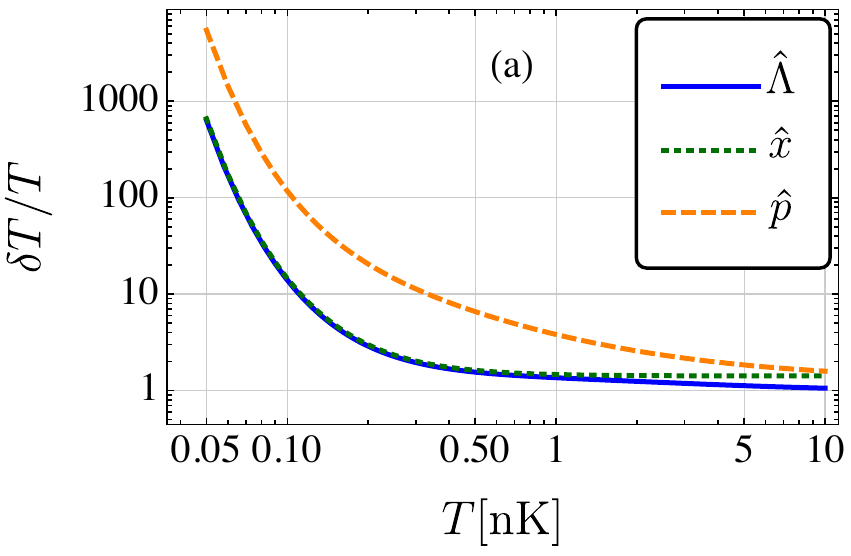}
		\includegraphics[width=0.68\columnwidth]{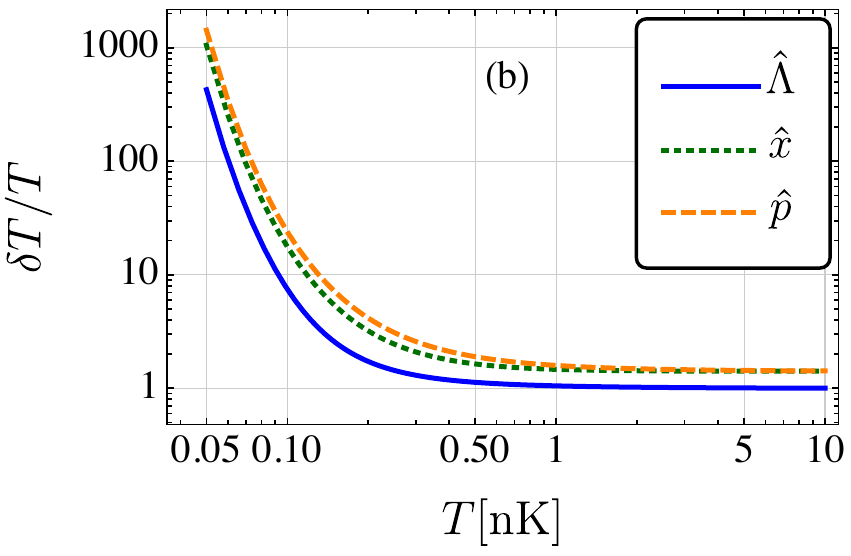}
		\includegraphics[width=0.68\columnwidth]{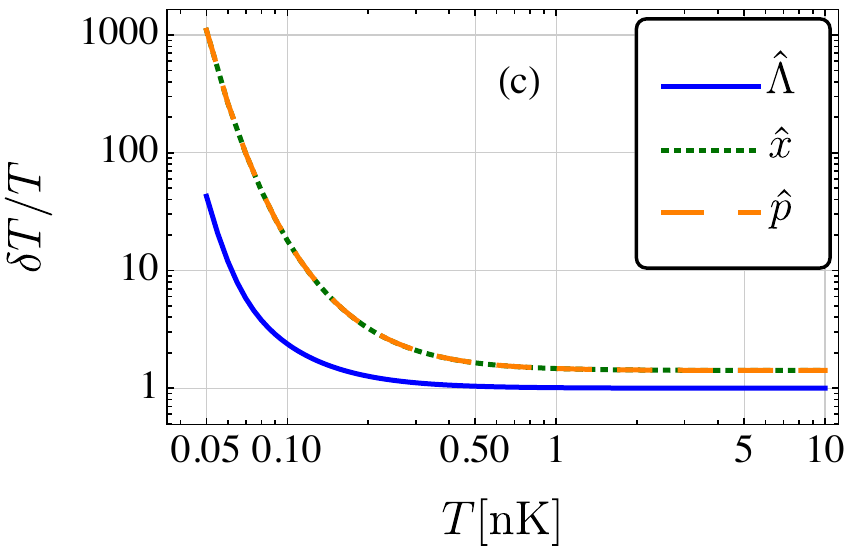}
		\caption{Same as in the upper panel of Fig. \ref{Therfig:fig4} of the main text, however temperature domain has been extended within the interval $ 0.05{\rm nK}\leq T \leq  10{\rm nK}$.}
		\label{AppB:Supfigb}
	\end{figure*}
In this appendix, we show the relative error for the low temperature domain down to ${\rm 50 pK}$. This reflects how much the relative error enhances as we move to lower values of the temperature. As such, the cases based on optimal measurements are shown in Fig. \ref{AppB:Supfiga}. Here the relative error of the $3d$ case starts deviating from the thermal profile at temperature below $T\simeq 60{\rm pK}$. Moreover, the relative errors based on the suboptimal measurements are shown in Figs. \ref{AppB:Supfigb} (a-c).
\end{document}